\documentclass[twocolumn, superscriptaddress]{revtex4-2} 
\usepackage[utf8]{inputenc}
\usepackage[a4paper, left=15mm, right=15mm, top=25mm, bottom=25mm]{geometry}
\usepackage{graphicx}
\usepackage{siunitx}
\usepackage{amssymb}
\usepackage{amsmath}
\usepackage{booktabs}
\usepackage{appendix}
\usepackage[dvipsnames]{xcolor}
\usepackage[normalem]{ulem}
\usepackage{braket}

\begin{document}

\title{Sub-barrier peaks in atom-atom-ion three-body recombination}
\date{September 3, 2026}

\author{Fabian Thielemann}
\affiliation{Physikalisches Institut, Albert-Ludwigs-Universität Freiburg, Hermann-Herder Str. 3, 79104
	Freiburg, Germany}
\affiliation{5.~Physikalisches Institut and Center for Integrated Quantum Science and Technology,
	Universität Stuttgart, Pfaffenwaldring 57, 70569 Stuttgart, Germany}
\author{Joachim Siemund}
\affiliation{Physikalisches Institut, Albert-Ludwigs-Universität Freiburg, Hermann-Herder Str. 3, 79104
	Freiburg, Germany}
\author{Tobias Schaetz}
\affiliation{Physikalisches Institut, Albert-Ludwigs-Universität Freiburg, Hermann-Herder Str. 3, 79104
	Freiburg, Germany}
\affiliation{EUCOR Centre for Quantum Science and Quantum Computing, Albert-Ludwigs-Universität Freiburg,
	79104 Freiburg, Germany}
\author{Krzysztof Jachymski}
\affiliation{Faculty of Physics, University of Warsaw, Pasteura 5, 02-093 Warsaw, Poland}

\begin{abstract}
	Hybrid atom-ion experiments have recently entered the few-partial wave regime in which magnetically tunable Feshbach resonances dominate the three-body loss dynamics. This necessitates a treatment of three-body recombination that can incorporate the role of the resonant dimer channel. Here, we develop a two-step Lindemann-type model for resonant atom-atom-ion three-body recombination that combines multichannel quantum-defect theory for the long-range polarization potential with a universal quantum-capture treatment of the inelastic atom-dimer step. We show that within our model, rate competition can lead to a peak in the three-body-recombination cross section of higher partial-wave channels at energies well below their corresponding centrifugal barrier. Investigating the resulting loss rates under numerically obtained non-thermal collision-energy distributions, we find that these effects survive the energy averaging and can be observed in state-of-the-art hybrid experimental setups. Our findings pave the way for a detailed characterization of atom-ion Feshbach resonances via the analysis of inelastic three-body-recombination processes.
\end{abstract}

\maketitle

\section{Introduction}

Ultracold collisions are governed by the quantum mechanical properties of matter. When
the wavelength of the two-particle wave function is on the order of the effective range of the
scattering potential, resonances with molecular bound states can drastically alter the collisional dynamics.
Examples of such phenomena include shape resonances~\cite{Henson2012,Jankunas2015}, where a metastable state resides in the open
channel, or Feshbach resonances, where coupling to a molecular bound state in a closed channel occurs.
The latter have become an important experimental tool, that allows  one to tune interactions between particles from weak to strong and repulsive to
attractive, and thus have been studied experimentally in a multitude of systems like homo- and heteronuclear
atom-atom~\cite{inouyeObservationFeshbachResonances1998,inouyeObservationHeteronuclearFeshbach2004},
atom-molecule~\cite{yangObservationMagneticallyTunable2019},
atom-ion~\cite{weckesserObservationFeshbachResonances2021} or molecule-molecule
combinations~\cite{parkFeshbachResonanceCollisions2023,Su2025}.

A strong enhancement of three-body losses near such resonances allows experimentalists to locate them~\cite{chin2010feshbach}.
However, from the theoretical perspective the three-particle system is rather complex and it can become challenging to describe and interpret experimental data~\cite{Esry1999,Petrov2003,Braaten2006}. In addition to two-body physics, genuine three-body effects such as Efimov resonances can occur, leading to additional features~\cite{Kraemer2006,Knoop2009}. The combination of quantum statistics, dependence on the mass ratio, and possible multichannel effects makes the problem extremely rich, such that a variety of approximate treatments can shed light on the underlying physics.

The case of atom-ion systems is rather special. Currently available experimental platforms involve just a single ion, enabling the detection of inelastic loss on the single-particle level~\cite{katz2022quantum}. Yet in these hybrid setups, the collision energy in combination with the long-range potential is large enough to involve multiple partial waves, such that most experiments operate in the semiclassical regime~\cite{Krukow2016}, where classical trajectory calculations can accurately describe the recombination rate~\cite{perez2015communication}. However, recent results have shown that at low energies there is a need for quantum mechanical treatment~\cite{feldker2020buffer,weckesserObservationFeshbachResonances2021,Thielemann2025,Siemund2026}. For ultracold atom-ion systems, recombination near a narrow Feshbach resonance involves a long-lived state and can be described using a similar two-step model to that established for neutral collisions~\cite{Beaufils2009,Li2018,Fouche2019,lecomteLossFeaturesUltracold2024,Liu2026}, which results in a Breit-Wigner loss profile. An additional subtlety that arises in hybrid traps is the non-thermal stationary state, in which the ion is driven by external electromagnetic fields and can have a much higher kinetic energy than the atoms in the gas~\cite{meir2016dynamics}.

In this work, we discuss in detail the application of the two-step recombination model to describe the experimental findings of hybrid atom-ion setups. We study in particular the lineshapes of higher-partial-wave resonances and the effect of the non-thermal energy distribution. Taking into account the competition between the dimer formation rate and the inelastic loss rate, we find that for resonances in open channels with angular momentum $\ell>0$ the scattering cross section can peak significantly below the height of the corresponding partial-wave barrier $E_\ell$. While presenting the results, we focus on the particular case of Li-Ba$^+$ with parameters similar to experimental conditions~\cite{Thielemann2025}, but our observations are general and can be applied to other hybrid atom-ion systems.

\section{Scattering theory}

\subsection{MQDT functions}

\begin{figure}[t]
	\centering
	\includegraphics[]{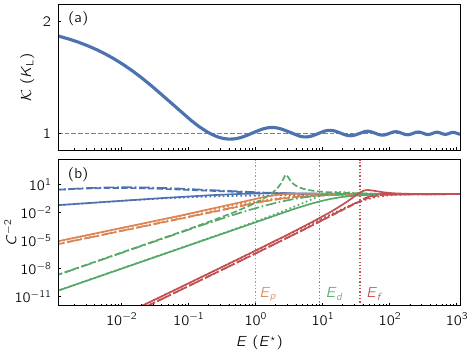}
	\caption{\textbf{Collision-energy scaling of inelastic-loss- and dimer-coupling rate.} (a) We show the inelastic
		atom-dimer collision parameter $\mathcal{K}$ in dependence on the collision energy $E$ in units of the
		Li-Ba$^+$ $s$-wave limit $E^\star$. At large energies it approaches the Langevin rate parameter $K_\text{L}$
		(dashed line), while at low energies it is enhanced by up to a factor of two. (b) We show the $C^{-2}$
		functions, which determine the energy scaling of the resonant dimer coupling rate, in dependence on $E$. The
		blue (orange/green/red) curves show the result for an open channel $s$-($p$/$d$/$f$-)wave resonance. We show
		the results for different background scattering lengths $a\in\{1, 10, -1, -10\}\, R^\star$
		(solid/dashed/dotted/dash-dotted lines). The vertical dotted lines indicate the height of the respective
		partial wave barriers $E_\ell$.}
	\label{fig:fig1}
\end{figure}

The atom-ion interaction at long range stems from the charge-induced dipole potential and behaves as $-C_4/r^4$, with the strength $C_4$ determined by the atomic polarizability. Following standard convention, we define the characteristic range $R^\star=\sqrt{2\mu C_4/\hbar^2}$  and energy $E^\star=\hbar^2/(2\mu R^{\star 2})$ with $\mu$ being the reduced mass. The energy $E^\star$ corresponds to the height of the p-wave centrifugal barrier, meaning that it can be treated as an estimate of the temperature needed to reach the $s$-wave limit and quantum degeneracy. The Schr\"{o}dinger equation for such a potential can be
solved analytically, provided that a proper short-range boundary condition is imposed~
\cite{vogtScatteringIonsPolarization1954}. This is done by mapping onto the Mathieu equation. The two linearly independent
solutions behave at short range like $f(r)\sim r \sin(1/r+\phi)$ and $g(r)\sim r \cos(1/r+\phi)$, where
$\phi$ is the short-range phase that connects to the scattering length via $a=R^\star\cot\phi$. Higher-order terms in the interaction potential mainly affect the short range physics, which is accounted for by the phase $\phi$. This simplification works especially well at low collision energies. The multi-channel quantum-defect theory (MQDT)
functions $C(E,\ell),\, \lambda(E,\ell)$ relate the short range solutions to the asymptotic scattering wave
functions and for the polarization potential can be calculated
analytically~\cite{idziaszekMultichannelQuantumdefectTheory2011}.
Together with the short range phase, or in the multichannel case the quantum defect matrix, they contain all
information about scattering properties. For example, the scattering phase shift in the presence of an
isolated Feshbach resonance at energy $E_0$ can be obtained as~\cite{miesMultichannelQuantumDefect1984}
\begin{equation}
	\xi(E,\ell)-\xi_{\rm bg}(E,\ell)=\tan^{-1}\left(\frac{\hbar\Gamma_m C^{-2}(E,\ell)}{E-E_0+\hbar\Gamma_m \tan\lambda(E,\ell)}\right)
	\label{eq:2body}
\end{equation}
with $\xi_{\rm bg}$ being the background phase shift of the open channel and the coupling strength between the two channels is described by $\Gamma_m$. It can be seen that the $C^{-2}$ function relates the bare coupling to the observed energy-dependent resonance width and accounts for the Wigner threshold laws, while the $\tan\lambda$ function merely provides the shift of the resonance from its bare value. The energy dependence of the $C^{-2}$ for the atom-ion potential is illustrated in Fig.~\ref{fig:fig1}b for several values of the scattering length, which is the only free parameter here.

\subsection{Reactive collisions}
The secondary process that we will need in order to characterize the three-body recombination is an inelastic atom-dimer collision involving an atom and a molecular ion. This process is assumed to be highly reactive and
thus very efficient, so one can treat it via a quantum capture model. In short, we assume absorbing
short-range boundary conditions for the atom-dimer scattering problem which results in unit loss probability
whenever the particles reach the reaction region. This results in a simple universal reactive rate constant
\cite{jachymskiQuantumTheoryReactive2013} controlled solely by the long-range potential. At high energies it approaches the well-known Langevin rate
\[
	\mathcal{K}_L=2\pi\sqrt{2C_4/\mu}\, ,
\]
while at very low energy, threshold effects lead to its quantum enhancement by a factor of 2. The rate constant changes at energies far below $E^\star$, making direct observation of this effect very challenging~\cite{Hoveler2022}. Due to the boundary condition which does not allow for the outgoing probability flux, there are no shape resonances in the reaction rate, and no free parameters involved. This assumption can be relaxed by adjusting the short-range behavior to be partially reflective~\cite{jachymskiQuantumTheoryReactive2013}.

\section{Two-step model}
Within the Lindemann mechanism, the unimolecular reaction is broken down into two steps: activation of the
molecule in a collision, and subsequent rearrangement. Ultracold three-body recombination can be approximated
by a similar two-step process
\cite{yurovskyThreebodyLossTrapped2003,waseemQuantitativeAnalysisWave2019,lecomteLossFeaturesUltracold2024}:
a resonant dimer formation, and a further collision which leads to a deeply
bound state. Here one proceeds with the scattering calculation assuming an isolated closed channel (the
collision complex), and many open channels corresponding to reaction
products~\cite{mottTheoryAtomicCollisions1965}. Taking into account only processes that involve the collision complex as an intermediate state, the atom-atom-ion recombination cross section is given by
\begin{align}
	\label{eq:scattering_cross_section}
	\sigma(k) = \frac{\pi}{k^2} \frac{\Gamma(k)\mathcal{K}(k)n}{(E-E_0)^2 / \hbar^2 +
		\left(\Gamma(k) + \mathcal{K}(k)n\right)^2/4},
\end{align}

where $E$ and $k=
	\sqrt{\frac{2\mu E}{\hbar^2}}$ are the collision energy and wavenumber, $E_0$ is the resonance position, $\Gamma(k)$ is the coupling rate to the dimer channel, $\mathcal{K}(k)$ is the inelastic reaction rate parameter and $n$ is the density of the atomic cloud at the position of the ion. We can express $\Gamma(k)=\Gamma_m C^{-2}(k,\ell)$ in terms of the short-range coupling $\Gamma_m$ and the energy-dependent MQDT function $C^{-2}(k,\ell)$, which follows threshold scaling laws at low energies and goes to unity at energies above the threshold $E_\ell$. Furthermore, the dimer relaxation rate $\mathcal{K}$ can be approximated by the Langevin rate. In Fig.~\ref{fig:fig1} we present $\mathcal{K}(k)$ and $C^{-2}(k)$ resulting from an MQDT calculation in reduced units. The presence of metastable bound states leads to shape resonances for $\ell>0$, witnessed by peaks of $C^{-2}(k,\ell)$ at energies below $E_\ell$.

\begin{figure}[t]
	\centering
	\includegraphics[]{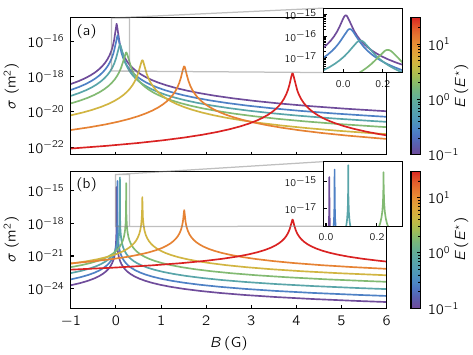}
	\caption{\textbf{Collision-energy dependence of the total three-body recombination cross section.} We show
		the total three-body recombination cross section $\sigma$ in dependence on the magnetic field $B$ for an
		open-channel $s$-wave (a) and $d$-wave resonance (b) based on Eq.~\ref{eq:scattering_cross_section}. The colors indicate
		singular collision energies ranging logarithmically from $0.1\,E^\star$ to $30\,E^\star$.
		In both cases we choose a background scattering length of $a=R^\star$, a short range coupling of
		$\Gamma_\text{m}=2\pi\times \SI{100}{\kilo\hertz}$, and atomic density of $n=\SI{1e17}{\per\cubic\meter}$. }
	\label{fig:fig2}
\end{figure}

\begin{figure}[t]
	\centering
	\includegraphics[]{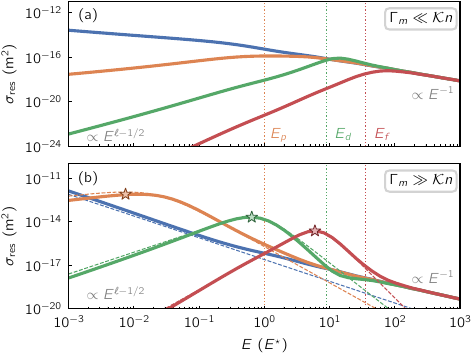}
	\caption{\textbf{Sub-barrier maxima of different partial-wave recombination resonances.} We show
	the resonant three-body recombination cross section $\sigma_\text{res}$ of the lowest four partial wave channels (blue/orange/green/red for $\ell={0,1,2,3}$) in dependence on the energy $E$. We choose $\Gamma_m = 2\pi\times \SI{1}{\hertz}$ to illustrate the loss-dominated (a), and $\Gamma_m=2\pi\times\SI{100}{\kilo\hertz}$ for the resonance-dominated regime (b).
	The vertical lines indicate the height of the corresponding partial
	wave barriers $E_\ell$. The dashed lines in (b) depict the threshold scaling approximation given in
	Eq.~\ref{eq:scattering_cross_section_low_energy} with the stars indicating the corresponding maxima given by
	Eq.~\ref{eq:k_peak}. In both (a) and (b) we use an atomic density of $n=\SI{1e17}{\per\cubic\meter}$.}
	\label{fig:fig3}
\end{figure}

At collision energies below the respective partial wave barrier $E< E_\ell$, and neglecting resonant contributions, the $C^{-2}$ functions follow
the threshold scaling law
\begin{align}
	C^{-2}(k, \ell) \approx A_\ell k^{2\ell+1},
\end{align}
where $A_\ell$ takes into account the different partial wave barriers. Considering that the $C^{-2}$ tends to unity
\begin{align}
	C^{-2} \xrightarrow{E> E_\ell} 1
\end{align}
we can approximate the partial wave dependent amplitudes as
\begin{align}
	A_\ell \approx \left( \frac{1}{k_\ell} \right)^{2\ell+1}
\end{align}
with $k_\ell=\sqrt{\frac{2\mu E_\ell}{\hbar^2}}$.
With these approximations Eq.~\eqref{eq:scattering_cross_section}, in the low energy regime, takes the form
\begin{align}
	\label{eq:scattering_cross_section_low_energy}
	\sigma(k) = \frac{\pi}{k^2} \frac{\Gamma_m A_\ell k^{2\ell+1} \mathcal{K}(k)n}{(E-E_0)^2 /\hbar^2+
		\left(\Gamma_m A_\ell k^{2\ell+1} + \mathcal{K}(k)n\right)^2/4}.
\end{align}
In the limit of $E\ll E_\ell$ the dimer coupling rate becomes small and we neglect variations in the inelastic loss rate, $\frac{\partial \mathcal{K}}{\partial k}\ll\frac{\partial C^{-2}}{\partial k}$ and $\mathcal{K}(k)n \approx \mathcal{K}_L n = \Gamma_d$ is constant. Then, the resonant scattering cross section at $E=E_0$ scales as
\begin{align}
	\label{eq:sigma_threshold_scaling}
	\sigma_\text{res} \xrightarrow{E\rightarrow0} \frac{4\pi}{k^2}\frac{\Gamma_m A_\ell k^{2\ell+1}}{\Gamma_d}\propto k^{2\ell-1}
\end{align}
Interestingly, in this form and for $\ell>0$ we find that $\sigma_\text{res}$ peaks at
\begin{align}
	\label{eq:k_peak}
	k_{\text{peak}} = \left(\frac{2\ell-1}{2\ell+3}\frac{\Gamma_d}{A_\ell \Gamma_m} \right)^{\frac{1}{2\ell+1}},
\end{align}
meaning that in the resonance-dominated regime, where ($\Gamma_m\gg\Gamma_d$), the resonant scattering cross section
$\sigma_\text{res}$ is maximal at energies well below $E_\ell$.
Intuitively, this is a result of the two-step nature of our model: for large dimer coupling rates $\Gamma_m$ the inelastic decay rate $\Gamma_d$ limits the fraction in Eq.~\ref{eq:scattering_cross_section_low_energy} and the universal prefactor $1/k^2$ determines the energy scaling.

Similarly, at energies above the threshold $E\gg E_\ell$, $\Gamma_d$ and $C^{-2}$ both become constant and the universal scaling
\begin{align}
	\label{eq:sigma_universal}
	\sigma_\text{res} \xrightarrow{E\rightarrow\infty} \frac{4\pi}{k^2}\frac{\Gamma_m \Gamma_d}{(\Gamma_m + \Gamma_d)^2} \propto k^{-2}
\end{align}
is recovered for all partial-wave channels.

\section{Energy dependence of three-body recombination}

To illustrate the distinct behavior of different partial wave resonances, we model a Li-Ba$^+$ $s$- and $d$-wave resonance with $\Gamma_m=2\pi\times\SI{100}{kHz}$ at a Li density of $n=\SI{1e17}{\meter^{-3}}$.  At these densities the inelastic reaction rate varies between $
	\mathcal{K}(0) \,n \approx\times\SI{960}{\per\second}$ at low and $\mathcal{K}(\infty)\, n \approx \SI{480}{\per\second}$ at large collision energies, meaning that at threshold the dimer coupling rate dominates. We relate the model to experimental observables, where typically magnetic Feshbach resonances are used to tune the scatterers to resonance, with $E_0 = \mu_\text{rel} (B-B_\text{res})$, with the relative magnetic moment $\mu_\text{rel}\approx \mu_\text{B}$, which is realistic for resonances stemming from coupling between spin singlet and triplet channels. For both partial waves we compute $\sigma (B)$  for logarithmically spaced singular collision energies between $0.1\,E^\star$ and $30\,E^\star$ using the full analytic solutions of $C^{-2}(k,\ell)$ for $a=R^\star$ and $\mathcal{K}(k)$ (see Fig.~\ref{fig:fig2}). Both resonances are similarly shifted to larger $B$ with increasing collision energy. While the $s$-wave resonance monotonically decreases in $\sigma_\text{res}$, the $d$-wave resonance initially grows in strength, peaks for $ E\approx E^\star$ before weakening and broadening at the same time. This characteristic peak at energies an order of magnitude below the $d$-wave limit is indicative of the resonance-dominated regime.

We now turn to investigate the resonant scattering cross section of the lowest four partial wave channels in the loss-dominated and the resonance-dominated regime. To this end, we calculate the energy dependent resonant scattering cross section $\sigma_\text{res}(E)$ from Eq.~\ref{eq:scattering_cross_section} for $E=E_0$, using the $C^{-2}$-functions obtained for $a=R^\star$ (see Fig.~\ref{fig:fig3}). We again use $n=\SI{1e17}{\meter^{-3}}$ while choosing $\Gamma_m=2\pi\times\SI{1}{Hz}$ to illustrate the loss-dominated and $\Gamma_m=2\pi\times\SI{100}{kHz}$ for the resonance-dominated regime. In the loss-dominated regime (Fig.~\ref{fig:fig3}a) and at energies $E\ll E_\ell$ the resonant scattering cross section follows the threshold scaling law of Eq.~\ref{eq:sigma_threshold_scaling}. This means that here $\sigma_\text{res}$ increases monotonically for all $\ell>0$. In the classical limit, $E\gg E_\ell$ all partial wave channels follow the universal decrease of Eq.~\ref{eq:sigma_universal}, resulting in a peak just above $E_\ell$ for all $\ell>0$.

In the resonance-dominated regime (Fig.~\ref{fig:fig3}b) we find  that all partial-wave channels with $\ell>0$ exhibit maxima fairly below their corresponding partial-wave barrier. At energies $E<E_\ell$ we find the resonant scattering cross section to be in good agreement with our approximation Eq.~\ref{eq:scattering_cross_section_low_energy}, exhibiting peaks agreeing with those predicted by Eq.~\ref{eq:k_peak}. Curiously, around the maximum location the peaking channel also dominates over all other partial waves, leading to a situation where e.g.\ the resonant scattering cross section of the $d$-wave channel has the strongest contribution to the total cross section at energies below the $s$-wave limit. Note that the effect found here is in contrast to the typical signature of shape resonances, which can increase the scattering cross section close to, but not orders of magnitude below, threshold.

\section{Non-thermal collision-energy averaged loss rates}

\begin{figure}[t]
	\centering
	\includegraphics[]{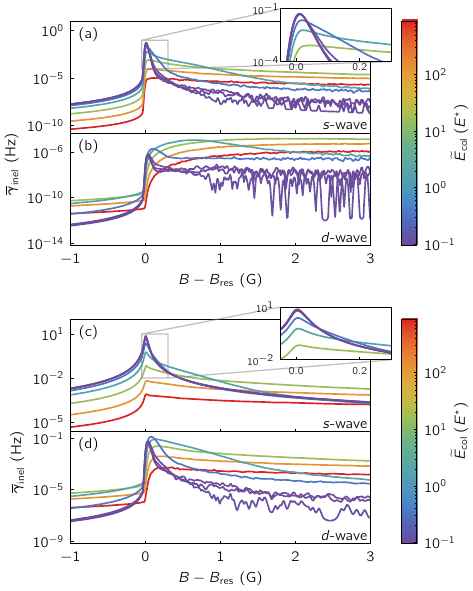}
	\caption{\textbf{Non-thermal collision-energy averaged line shape of the three-body recombination loss rate.} We show
		the dependence of the energy-averaged inelastic loss rate $\overline{\gamma}_\text{inel}$ (Eq.~\ref{eq:gamma_av}) on the magnetic
		field $B$. The collision energy is increased by applying an
		electric field that displaces the ion from the center of its rf-trap, resulting in a non-thermal energy distribution.
		Here we show the resulting lineshape for
		median collision energies $\widetilde{E}_\text{col}$ ranging from $0.1\,E^\star$ (blue) to
		$900\,E^\star$ (red). The behavior is shown for the loss-dominated regime with $\Gamma_m = 2\pi\times \SI{1}{\hertz}$ (a,b) and
		the resonance-dominated regime with $\Gamma_m=2\pi\times\SI{100}{\kilo\hertz}$ (c,d) for an open-channel $s$- (a, c) and $d$-wave (b, d)
		resonance respectively. The insets in (a) and (c) enlarge the region around the peak of the $s$-wave resonance.}
	\label{fig:fig4}
\end{figure}

In order to compare our model to experimental findings it is necessary to take the collision-energy distribution into account. In hybrid atom-ion experiments this has some peculiarities: as the ion is held in an rf-trap, in which a time-dependent potential creates a ponderomotive harmonic confinement, it never thermalizes with the atoms. It is thus necessary to handle the ionic and atomic contributions to the collision energy separately. Further, the driven nature of the ion's confinement leads to a non-thermal energy distribution with a power-law tail $P\propto E^{-\alpha-1}$ towards higher energies~\cite{meir2016dynamics}.

We obtain estimates for the equilibrium velocity distributions of a trapped ion in an ultracold atomic bath by numerical integration of the classical equations of motion, the details of which are found in Appendix~\ref{apdx:num_atom_ion}.
The collision energy is tuned by applying a dc displacement field, which increases the excess micromotion of the ion, while keeping the atomic bath temperature constant, as realized in~\cite{Thielemann2025}.
From the computed ion trajectories, we draw random ion velocities $\mathbf{v}_\text{i}$ to calculate the collision energy in the center-of-mass frame
\begin{align}
	E_\text{col} = \frac{1}{2} \mu (\mathbf{v}_\text{i} - \mathbf{v}_\text{a})^2,
\end{align}
where the atom velocities $\mathbf{v}_\text{a}$ are drawn from a thermal distribution. As the resulting collision-energy distributions do not possess a mean for $\alpha\leq2$, we characterize them by their numerical median
\begin{align}
	\widetilde{E}_\text{col} = \text{med}\left[E_{\text{col}}^i\right].
\end{align}
The typical experimental observable is the inelastic loss rate of the ion $\gamma_\text{inel} = \sigma n v$, which is measured as an exponential decrease of the ion survival probability $P\propto\exp\left\{-\gamma_\text{inel}\, t\right\}$. This is in contrast to three-body-recombination in cold neutral ensembles, where a super-exponential decay with $\dot{n}\propto -n^3$ occurs. We compute the non-thermal average of the inelastic-loss rate
\begin{align}
	\label{eq:gamma_av}
	\overline{\gamma}_\text{inel} = \frac{1}{N}\sum_{j=1}^N \sigma(k_{j})\, n \, v_{\text{rel},j},
\end{align}
where $v_{\text{rel},j}$ are the relative atom-ion speeds randomly sampled and $k_j$ is the corresponding wavenumber. This approach also accounts for the anisotropic collision-energy distributions that are encountered when micromotion along the axial degree of freedom is much smaller than in the radial plane.

For the same resonance parameters as were used in Fig.~\ref{fig:fig2}, we show the collision-energy averaged loss rates in Fig.~\ref{fig:fig4}. The first apparent effect is an asymmetric broadening of the loss feature, which is more pronounced for  resonances in the loss-dominated (Figs.~\ref{fig:fig4}a,b) than for those in the resonance-dominated regime (Figs.~\ref{fig:fig4}c,d), and is the result of the collision-energy distribution being bounded from below. Noticeably, we find that with experimentally realistic parameters, the qualitative difference between the $s$-wave and $d$-wave resonances persists in both regimes. While the strength of the $s$-wave resonance decreases monotonically, the $d$-wave resonance peaks at temperatures above the lower limit of $\widetilde{E}_\text{col}\approx 0.1\,E^\star$ obtained in our simulations. The shift of the resonance peak position in $B$ turns out to be much stronger for the $d$-wave resonance. For singular collision energies, according to Eq.~\ref{eq:scattering_cross_section}, the shift is predicted to be of the same magnitude, independent of $\ell$. However, when taking the collision-energy average, larger energies (corresponding to larger shifts) contribute less (more) to the total inelastic loss rate for the $s$-wave ($d$-wave) resonance, leading to a smaller (larger) effective shift.

\begin{figure}[t]
	\centering
	\includegraphics[]{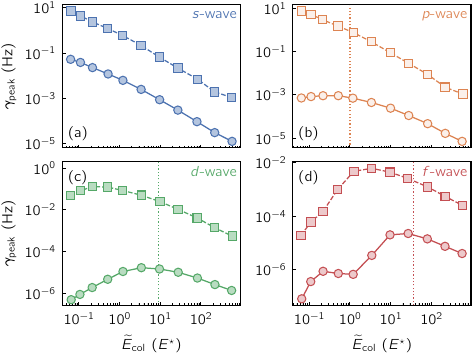}
	\caption{\textbf{Collision-energy averaged peak inelastic loss rates.} We show the peak loss rate
		$\gamma_\text{peak}$ of the collision-energy averaged resonances in dependence on the median collision energy
		$\widetilde{E}_\text{col}$ for the
		lowest four open-channel partial waves $\ell\in\{0,1,2,3\}$ (a-d). The loss rates are
		calculated in the loss-dominated ($\Gamma_m=2\pi\times\SI{1}{\hertz}$, circles) and resonance dominated
		($\Gamma_m=2\pi\times\SI{100}{\kilo\hertz}$, squares) regime and for $\widetilde{E}_\text{col}$ that are achievable in state-of-the-art hybrid experiments. In
		both cases we assume an atomic density of $n=\SI{1e17}{\per\cubic\meter}$ and that the collision energy is
		increased by displacing the ion radially from the center of its trap. The vertical dotted lines show the corresponding partial wave barriers.}
	\label{fig:fig5}
\end{figure}

Finally, we explore whether these characteristics can be investigated in state-of-the-art hybrid atom-ion experiments. To this end, we compute the peak inelastic loss rate $\gamma_\text{peak}=\max\left\{\overline{\gamma}_\text{inel}(B)\right\}$ for the lowest four partial wave channels (see Fig.~\ref{fig:fig5}). Here, we find that in the loss-dominated regime and within the range of experimentally accessible collision energies (and distributions) the peak loss rates qualitatively follow the behavior previously described for scattering cross sections at singular energies: all partial waves with $\ell>0$ exhibit a peak close to their partial wave barrier. In the resonance-dominated regime and for our choice of parameters, $s$- and $p$-wave resonances become almost indistinguishable with respect to their $\gamma_\text{peak}$ scaling. The maximum of the scattering cross section predicted by Eq.~\ref{eq:k_peak} for $\ell=1$ appears at lower energies and thus $\gamma_\text{peak}$ shows a decay which is similar to $\ell=0$. For higher-partial-wave channels such as $d$ and $f$, the maxima of $\gamma_\text{peak}$ appear well below the partial wave barrier and within the accessible energy range. This is qualitatively consistent with the experimental observation of higher-partial wave resonances below the $s$-wave limit described in~\cite{Thielemann2025}.

\section{Conclusion and outlook}
We have developed a Lindemann-type two-step model for the description of resonant atom-atom-ion recombination processes. In our model the first step consists of the resonant formation of a dimer in the vicinity of an atom-ion Feshbach resonance and its energy dependence is described in an MQDT framework. The second step, the collision with a second atom resulting in the formation of a deeply bound molecular state, is treated with a quantum capture model. This results in a Breit-Wigner expression for the scattering cross section in which the competition between the dimer-coupling and the reactive collision rate becomes apparent. In a regime where the close-range coupling rate is much larger than the reactive-collision rate this leads to a maximum in the scattering cross section at energies well below the partial wave barrier. We have derived an approximate expression for the energy at which this peak occurs and found it to be in good agreement with the full description. Averaging over numerically obtained collision-energy distributions we found these effects to persist under experimentally realistic conditions.

Our approach offers a simple and intuitive picture to understand the energy-dependence of atom-atom-ion recombination in the ultracold regime. However, it does not capture all aspects of scattering processes in this regime. It was shown that, for collisions at sufficiently low energies and in the case of a near-degenerate gas, the quantum statistics of the bath has a significant impact on the loss dynamics~\cite{Siemund2026}. While these effects can be phenomenologically included in the two-step model, e.g.\ by modifying the recombination rate $\mathcal{K}(k)$, their full description would require treating all three scatterers in a common frame, for instance via a multichannel three-body calculation in hyperspherical coordinates. Further, our model treats the resonance as being isolated with a well defined background scattering length. The Feshbach spectra of atom-ion systems were experimentally observed to be partially very densely populated. In this regime our individual treatment would break down, requiring a more general sum over resonant channels. Finally, in atom-ion experiments with a 1:1 mass-ratio, the occurrence of trap-induced bound states has been observed~\cite{pinkas2023trap}. These rapid re-collisions lead to an increase of the inelastic loss rate. Although for the trap parameters and mass ratios used in this manuscript the effect was estimated to be negligible, it could be incorporated into our model by introducing an additional, energy-dependent loss channel.

We expect our model to facilitate the experimental characterization of atom-ion Feshbach resonances, based on observables such as the energy dependence of the three-body-recombination loss rate, or collision-energy induced shifts of the resonance position. Here, the density dependence of the sub-barrier peak position can serve as an additional experimental handle.  Our model can be straightforwardly extended to denser spectra, different hybrid-trap configurations or other atomic species, and we thus believe it to be a useful tool for the quantitative analysis of forthcoming few-partial-wave atom-ion experiments.

\section*{Acknowledgments}
The authors thank Paul Julienne for fruitful discussions.
This project has received funding from the European Research Council (ERC) under the European Union's Horizon 2020 research and innovation program (Grant No. 648330), the Deutsche Forschungsgemeinschaft (DFG, Grant No. SCHA 973/9-1-3017959 and SCHA 973/10-1), and the Georg H. Endress Foundation. FT, JS, and TS acknowledge financial support from the DFG via the RTG DYNCAM 2717. FT acknowledges funding from the German Federal Ministry of Research, Technology and Space under the CiRQus grant. KJ was supported by the Polish National Agency for Academic Exchange (NAWA) via the Polish Returns 2019 program.

\section*{Data availability}
The data supporting the findings of this manuscript will be published on Zenodo along with publication.

\bibliography{manually_added}

@article{Hoveler2022,
  title = {Observation of quantum capture in an ion-molecule reaction},
  author = {H\"oveler, Katharina and Deiglmayr, Johannes and Agner, Josef A. and Hahn, Rapha\"el and Zhelyazkova, Valentina and Merkt, Fr\'ed\'eric},
  journal = {Phys. Rev. A},
  volume = {106},
  issue = {5},
  pages = {052806},
  numpages = {8},
  year = {2022},
  month = {Nov},
  publisher = {American Physical Society},
  doi = {10.1103/PhysRevA.106.052806},
  url = {https://link.aps.org/doi/10.1103/PhysRevA.106.052806}
}

@article{Su2025,
  title = {Broad {F}eshbach Resonance with a Large Background Scattering Length in a Fermionic Atom-Molecule Mixture},
  author = {Su, Zhen and Shou, Tong-Hui and Yang, Huan and Cao, Jin and Wang, Bo-Yuan and Xie, Ting and Rui, Jun and Zhao, Bo and Pan, Jian-Wei},
  journal = {Phys. Rev. Lett.},
  volume = {135},
  issue = {21},
  pages = {213401},
  numpages = {6},
  year = {2025},
  month = {Nov},
  publisher = {American Physical Society},
  doi = {10.1103/95nd-4tzv},
  url = {https://link.aps.org/doi/10.1103/95nd-4tzv}
}

@article{Liu2026,
  title={Orbital-resolved three-body recombination across a p-wave {F}eshbach resonance in ultracold $^6${L}i},
  author={Liu, Shaokun and Xu, Zhekang and Peng, Shuai and Peng, Sijia and Shu, Tangqian and Li, Jiaming and Luo, Le},
  journal={Rep. Prog. Phys.},
  volume={89},
  number={2},
  pages={020502},
  year={2026},
  publisher={IOP Publishing}
}

@article{Beaufils2009,
  title = {{F}eshbach resonance in $d$-wave collisions},
  author = {Beaufils, Q. and Crubellier, A. and Zanon, T. and Laburthe-Tolra, B. and Mar\'echal, E. and Vernac, L. and Gorceix, O.},
  journal = {Phys. Rev. A},
  volume = {79},
  issue = {3},
  pages = {032706},
  numpages = {8},
  year = {2009},
  month = {Mar},
  publisher = {American Physical Society},
  doi = {10.1103/PhysRevA.79.032706},
  url = {https://link.aps.org/doi/10.1103/PhysRevA.79.032706}
}

@article{Li2018,
  title = {Three-Body Recombination near a Narrow {F}eshbach Resonance in $^{6}\mathrm{Li}$},
  author = {Li, Jiaming and Liu, Ji and Luo, Le and Gao, Bo},
  journal = {Phys. Rev. Lett.},
  volume = {120},
  issue = {19},
  pages = {193402},
  numpages = {6},
  year = {2018},
  month = {May},
  publisher = {American Physical Society},
  doi = {10.1103/PhysRevLett.120.193402},
  url = {https://link.aps.org/doi/10.1103/PhysRevLett.120.193402}
}

@article{Fouche2019,
  title = {Quantitative analysis of losses close to a $d$-wave open-channel {F}eshbach resonance in $^{39}\mathrm{K}$},
  author = {Fouch\'e, L. and Boiss\'e, A. and Berthet, G. and Lepoutre, S. and Simoni, A. and Bourdel, T.},
  journal = {Phys. Rev. A},
  volume = {99},
  issue = {2},
  pages = {022701},
  numpages = {6},
  year = {2019},
  month = {Feb},
  publisher = {American Physical Society},
  doi = {10.1103/PhysRevA.99.022701},
  url = {https://link.aps.org/doi/10.1103/PhysRevA.99.022701}
}

@article{Siemund2026,
  title={Quantum statistics on atom-ion {F}eshbach resonances},
  author={Siemund, Joachim and Thielemann, Fabian and Grieshaber, Jonathan and Wu, Wei and Mullan, Patrick and Giannakeas, Panagiotis and Jachymski, Krzysztof and Schaetz, Tobias},
  journal={arXiv:2606.26995},
  year={2026}
}

@article{Thielemann2025,
  title = {Exploring Atom-Ion {F}eshbach Resonances below the $s$-Wave Limit},
  author = {Thielemann, Fabian and Siemund, Joachim and von Schoenfeld, Daniel and Wu, Wei and Weckesser, Pascal and Jachymski, Krzysztof and Walker, Thomas and Schaetz, Tobias},
  journal = {Phys. Rev. X},
  volume = {15},
  issue = {1},
  pages = {011051},
  numpages = {11},
  year = {2025},
  month = {Mar},
  publisher = {American Physical Society},
  doi = {10.1103/PhysRevX.15.011051},
  url = {https://link.aps.org/doi/10.1103/PhysRevX.15.011051}
}

@article{Henson2012,
  title={Observation of resonances in {P}enning ionization reactions at sub-kelvin temperatures in merged beams},
  author={Henson, Alon B and Gersten, Sasha and Shagam, Yuval and Narevicius, Julia and Narevicius, Edvardas},
  journal={Science},
  volume={338},
  number={6104},
  pages={234--238},
  year={2012},
  publisher={American Association for the Advancement of Science}
}

@article{Jankunas2015,
  title={Observation of orbiting resonances in {H}e (3{S}1)+ {NH}3 {P}enning ionization},
  author={Jankunas, Justin and Jachymski, Krzysztof and Hapka, Micha{\l} and Osterwalder, Andreas},
  journal={J. Chem. Phys.},
  volume={142},
  number={16},
  year={2015},
  publisher={AIP Publishing}
}

@article{Kraemer2006,
  title={Evidence for {E}fimov quantum states in an ultracold gas of caesium atoms},
  author={Kraemer, Tobias and Mark, Manfred and Waldburger, Philipp and Danzl, Johann G and Chin, Cheng and Engeser, Bastian and Lange, Almar D and Pilch, Karl and Jaakkola, Antti and N{\"a}gerl, H-C and others},
  journal={Nature},
  volume={440},
  number={7082},
  pages={315--318},
  year={2006},
  publisher={Nature Publishing Group UK London}
}

@article{Knoop2009,
  title={Observation of an {E}fimov-like trimer resonance in ultracold atom--dimer scattering},
  author={Knoop, S and Ferlaino, F and Mark, M and Berninger, M and Sch{\"o}bel, H and N{\"a}gerl, H-C and Grimm, R},
  journal={Nat. Phys.},
  volume={5},
  number={3},
  pages={227--230},
  year={2009},
  publisher={Nature Publishing Group UK London}
}

@article{Petrov2003,
  title = {Three-body problem in {F}ermi gases with short-range interparticle interaction},
  author = {Petrov, D. S.},
  journal = {Phys. Rev. A},
  volume = {67},
  issue = {1},
  pages = {010703(R)},
  numpages = {4},
  year = {2003},
  month = {Jan},
  publisher = {American Physical Society},
  doi = {10.1103/PhysRevA.67.010703},
  url = {https://link.aps.org/doi/10.1103/PhysRevA.67.010703}
}

@article{Esry1999,
  title = {Recombination of Three Atoms in the Ultracold Limit},
  author = {Esry, B. D. and Greene, Chris H. and Burke, James P.},
  journal = {Phys. Rev. Lett.},
  volume = {83},
  issue = {9},
  pages = {1751--1754},
  numpages = {0},
  year = {1999},
  month = {Aug},
  publisher = {American Physical Society},
  doi = {10.1103/PhysRevLett.83.1751},
  url = {https://link.aps.org/doi/10.1103/PhysRevLett.83.1751}
}

@article{Braaten2006,
  title={Universality in few-body systems with large scattering length},
  author={Braaten, Eric and Hammer, H-W},
  journal={Phys. Rep.},
  volume={428},
  number={5-6},
  pages={259--390},
  year={2006},
  publisher={Elsevier}
}

@article{Krukow2016,
  title = {Energy Scaling of Cold Atom-Atom-Ion Three-Body Recombination},
  author = {Kr\"ukow, Artjom and Mohammadi, Amir and H\"arter, Arne and Denschlag, Johannes Hecker and P\'erez-R\'{\i}os, Jes\'us and Greene, Chris H.},
  journal = {Phys. Rev. Lett.},
  volume = {116},
  issue = {19},
  pages = {193201},
  numpages = {5},
  year = {2016},
  month = {May},
  publisher = {American Physical Society},
  doi = {10.1103/PhysRevLett.116.193201},
  url = {https://link.aps.org/doi/10.1103/PhysRevLett.116.193201}
}

@article{perez2015communication,
  title={Communication: Classical threshold law for ion-neutral-neutral three-body recombination},
  author={P{\'e}rez-R{\'\i}os, Jes{\'u}s and Greene, Chris H},
  journal={J. Chem. Phys.},
  volume={143},
  number={4},
  year={2015},
  publisher={AIP Publishing}
}

@article{Rackauckas2017,
 author = {Rackauckas, Christopher and Nie, Qing},
 doi = {10.5334/jors.151},
 journal = {Journal of Open Research Software},
 month = {May},
 title = {DifferentialEquations.jl – A Performant and Feature-Rich Ecosystem for Solving Differential Equations in Julia},
 year = {2017}
}

@article{Fuerst2018,
author = {F{\"u}rst, H A and Ewald, N V and Secker, T and Joger, J and Feldker, T and Gerritsma, R},
title = {Prospects of reaching the quantum regime in {L}i–{Y}b+ mixtures},
journal = {J. Phys. B: At. Mol. Opt. Phys.},
doi = {10.1088/1361-6455/aadd7d},
url = {https://doi.org/10.1088/1361-6455/aadd7d},
year = {2018},
month = {sep},
publisher = {IOP Publishing},
volume = {51},
number = {19},
pages = {195001},
}

@article{leibfried2003quantum,
  title={Quantum dynamics of single trapped ions},
  author={Leibfried, Dietrich and Blatt, Rainer and Monroe, Christopher and Wineland, David},
  journal={Rev. Mod. Phys.},
  volume={75},
  number={1},
  pages={281},
  year={2003},
  publisher={APS}
}

@article{meir2016dynamics,
  title={Dynamics of a ground-state cooled ion colliding with ultracold atoms},
  author={Meir, Ziv and Sikorsky, Tomas and Ben-Shlomi, Ruti and Akerman, Nitzan and Dallal, Yehonatan and Ozeri, Roee},
  journal={Phys. Rev. Lett.},
  volume={117},
  number={24},
  pages={243401},
  year={2016},
  publisher={APS}
}

@article{idziaszekMultichannelQuantumdefectTheory2011,
  title = {Multichannel Quantum-Defect Theory for Ultracold Atom–Ion Collisions},
  author = {Idziaszek, Zbigniew and Simoni, Andrea and Calarco, Tommaso and Julienne, Paul S},
  date = {2011-08-08},
  year = {2011},
  journal = {New J. Phys.},
  shortjournal = {New J. Phys.},
  volume = {13},
  number = {8},
  pages = {083005},
  issn = {1367-2630},
  doi = {10.1088/1367-2630/13/8/083005},
  url = {https://iopscience.iop.org/article/10.1088/1367-2630/13/8/083005},
  urldate = {2024-12-09},
  langid = {english},
}

@article{inouyeObservationFeshbachResonances1998,
  title = {Observation of {{{F}eshbach}} Resonances in a {{Bose}}–{{Einstein}} Condensate},
  author = {Inouye, S. and Andrews, M. R. and Stenger, J. and Miesner, H.-J. and Stamper-Kurn, D. M. and Ketterle, W.},
  date = {1998-03},
  year = {1998},
  journal = {Nature},
  volume = {392},
  number = {6672},
  pages = {151--154},
  publisher = {Nature Publishing Group},
  issn = {1476-4687},
  doi = {10.1038/32354},
  url = {https://www.nature.com/articles/32354},
  urldate = {2024-08-30},
  langid = {english},
}

@article{inouyeObservationHeteronuclearFeshbach2004,
  title = {Observation of {{Heteronuclear {F}eshbach Resonances}} in a {{Mixture}} of {{Bosons}} and {{Fermions}}},
  author = {Inouye, S. and Goldwin, J. and Olsen, M. L. and Ticknor, C. and Bohn, J. L. and Jin, D. S.},
  date = {2004-10-25},
  year = {2004},
  journal = {Phys. Rev. Lett.},
  shortjournal = {Phys. Rev. Lett.},
  volume = {93},
  number = {18},
  pages = {183201},
  publisher = {American Physical Society},
  doi = {10.1103/PhysRevLett.93.183201},
  url = {https://link.aps.org/doi/10.1103/PhysRevLett.93.183201},
  urldate = {2024-09-05},
}

@article{jachymskiQuantumTheoryReactive2013,
  title = {Quantum {{Theory}} of {{Reactive Collisions}} for $1 / r^n$ {{Potentials}}},
  author = {Jachymski, Krzysztof and Krych, Michał and Julienne, Paul S. and Idziaszek, Zbigniew},
  date = {2013-05-23},
  year = {2013},
  journal = {Phys. Rev. Lett.},
  shortjournal = {Phys. Rev. Lett.},
  volume = {110},
  number = {21},
  pages = {213202},
  issn = {0031-9007, 1079-7114},
  doi = {10.1103/PhysRevLett.110.213202},
  url = {https://link.aps.org/doi/10.1103/PhysRevLett.110.213202},
  urldate = {2024-03-06},
  langid = {english},
}

@article{lecomteLossFeaturesUltracold2024,
  title = {Loss Features in Ultracold $^{162}${{Dy}} Gases: {{Two-}} versus Three-Body-Processes},
  shorttitle = {Loss Features in Ultracold \$\textasciicircum\{162\}\textbackslash mathrm\{\vphantom\}{{Dy}}\vphantom\{\}\$ Gases},
  author = {Lecomte, Maxime and Journeaux, Alexandre and Renaud, Loan and Dalibard, Jean and Lopes, Raphael},
  date = {2024-02-14},
  year = {2024},
  journal = {Phys. Rev. A},
  shortjournal = {Phys. Rev. A},
  volume = {109},
  number = {2},
  pages = {023319},
  publisher = {American Physical Society},
  doi = {10.1103/PhysRevA.109.023319},
  url = {https://link.aps.org/doi/10.1103/PhysRevA.109.023319},
  urldate = {2024-03-07},
}

@article{miesMultichannelQuantumDefect1984,
  title = {A Multichannel Quantum Defect Analysis of Two‐state Couplings in Diatomic Molecules},
  author = {Mies, Frederick H. and Julienne, Paul S.},
  date = {1984-03-15},
  year = {1984},
  journal = {J. Chem. Phys.},
  shortjournal = {J. Chem. Phys.},
  volume = {80},
  number = {6},
  pages = {2526--2536},
  issn = {0021-9606},
  doi = {10.1063/1.447046},
  url = {https://doi.org/10.1063/1.447046},
  urldate = {2025-08-05},
}

@book{mottTheoryAtomicCollisions1965,
  title = {The {{Theory}} of {{Atomic Collisions}}},
  author = {Mott, Sir Nevill Francis and Massey, Sir Harrie Stewart Wilson},
  date = {1965},
  year = {1965},
  eprinttype = {googlebooks},
  publisher = {Clarendon Press},
  isbn = {978-0-19-851242-4},
  langid = {english},
  pagetotal = {890}
}

@article{parkFeshbachResonanceCollisions2023,
  title = {A {{{F}eshbach}} Resonance in Collisions between Triplet Ground-State Molecules},
  author = {Park, Juliana J. and Lu, Yu-Kun and Jamison, Alan O. and Tscherbul, Timur V. and Ketterle, Wolfgang},
  date = {2023-02},
  year = {2023},
  journal = {Nature},
  volume = {614},
  number = {7946},
  pages = {54--58},
  publisher = {Nature Publishing Group},
  issn = {1476-4687},
  doi = {10.1038/s41586-022-05635-8},
  url = {https://www.nature.com/articles/s41586-022-05635-8},
  urldate = {2024-04-09},
  langid = {english},
}

@article{vogtScatteringIonsPolarization1954,
  title = {Scattering of {{Ions}} by {{Polarization Forces}}},
  author = {Vogt, Erich and Wannier, Gregory H.},
  date = {1954-09-01},
  year = {1954},
  journal = {Phys. Rev.},
  shortjournal = {Phys. Rev.},
  volume = {95},
  number = {5},
  pages = {1190--1198},
  publisher = {American Physical Society},
  doi = {10.1103/PhysRev.95.1190},
  url = {https://link.aps.org/doi/10.1103/PhysRev.95.1190},
  urldate = {2025-08-05},
}

@article{waseemQuantitativeAnalysisWave2019,
  title = {Quantitative Analysis of p-Wave Three-Body Losses via a Cascade Process},
  author = {Waseem, Muhammad and Yoshida, Jun and Saito, Taketo and Mukaiyama, Takashi},
  date = {2019-05-28},
  year = {2019},
  journal = {Phys. Rev. A},
  shortjournal = {Phys. Rev. A},
  volume = {99},
  number = {5},
  pages = {052704},
  issn = {2469-9926, 2469-9934},
  doi = {10.1103/PhysRevA.99.052704},
  url = {https://link.aps.org/doi/10.1103/PhysRevA.99.052704},
  urldate = {2024-03-06},
  langid = {english},
}

@article{weckesserObservationFeshbachResonances2021,
  title = {Observation of {{{F}eshbach}} Resonances between a Single Ion and Ultracold Atoms},
  author = {Weckesser, Pascal and Thielemann, Fabian and Wiater, Dariusz and Wojciechowska, Agata and Karpa, Leon and Jachymski, Krzysztof and Tomza, Michał and Walker, Thomas and Schaetz, Tobias},
  date = {2021-12},
  year = {2021},
  journal = {Nature},
  volume = {600},
  number = {7889},
  pages = {429--433},
  publisher = {Nature Publishing Group},
  issn = {1476-4687},
  doi = {10.1038/s41586-021-04112-y},
  url = {https://www.nature.com/articles/s41586-021-04112-y},
  urldate = {2024-03-06},
  langid = {english},
}

@article{yangObservationMagneticallyTunable2019,
  title = {Observation of Magnetically Tunable {{{F}eshbach}} Resonances in Ultracold $^{23}${Na} $^{40}${K} + $^{40}${K} Collisions},
  author = {Yang, Huan and Zhang, De-Chao and Liu, Lan and Liu, Ya-Xiong and Nan, Jue and Zhao, Bo and Pan, Jian-Wei},
  date = {2019-01-18},
  year = {2019},
  journal = {Science},
  shortjournal = {Science},
  volume = {363},
  number = {6424},
  pages = {261--264},
  issn = {0036-8075, 1095-9203},
  doi = {10.1126/science.aau5322},
  url = {https://www.science.org/doi/10.1126/science.aau5322},
  urldate = {2024-06-05},
  langid = {english},
}

@article{yurovskyThreebodyLossTrapped2003,
  title = {Three-Body Loss of Trapped Ultracold 87 {{Rb}} Atoms Due to a {{{F}eshbach}} Resonance},
  author = {Yurovsky, V. A. and Ben-Reuven, A.},
  date = {2003-05-13},
  year = {2003},
  journal = {Phys. Rev. A},
  shortjournal = {Phys. Rev. A},
  volume = {67},
  number = {5},
  pages = {050701},
  issn = {1050-2947, 1094-1622},
  doi = {10.1103/PhysRevA.67.050701},
  url = {https://link.aps.org/doi/10.1103/PhysRevA.67.050701},
  urldate = {2024-07-30},
  langid = {english},
}

@article{feldker2020buffer,
  title={Buffer gas cooling of a trapped ion to the quantum regime},
  author={Feldker, T and F{\"u}rst, H and Hirzler, H and Ewald, NV and Mazzanti, M and Wiater, D and Tomza, M and Gerritsma, R},
  journal={Nat. Phys.},
  volume={16},
  number={4},
  pages={413--416},
  year={2020},
  publisher={Nature Publishing Group UK London}
}

@article{chin2010Feshbach,
  title={{F}eshbach resonances in ultracold gases},
  author={Chin, Cheng and Grimm, Rudolf and Julienne, Paul and Tiesinga, Eite},
  journal={Rev. Mod. Phys.},
  volume={82},
  number={2},
  pages={1225--1286},
  year={2010},
  publisher={APS}
}

@article{pinkas2023trap,
  title={Trap-assisted formation of atom--ion bound states},
  author={Pinkas, Meirav and Katz, Or and Wengrowicz, Jonathan and Akerman, Nitzan and Ozeri, Roee},
  journal={Nat. Phys.},
  volume={19},
  number={11},
  pages={1573--1578},
  year={2023},
  publisher={Nature Publishing Group UK London}
}

@article{katz2022quantum,
  title={Quantum logic detection of collisions between single atom--ion pairs},
  author={Katz, Or and Pinkas, Meirav and Akerman, Nitzan and Ozeri, Roee},
  journal={Nature physics},
  volume={18},
  number={5},
  pages={533--537},
  year={2022},
  publisher={Nature Publishing Group UK London}
}

\appendix

\section{Numerical atom-ion simulation}
\label{apdx:num_atom_ion}

To obtain numerical estimates of the collision energy distribution, we integrate the classical equations of motion (eom) of a single trapped ion, under collisions with free ultracold atoms. The total force acting on the ion is
\begin{align}
	\mathbf{F}_\text{ion} = \mathbf{F}_\text{trap} + \mathbf{F}_\text{dc} + \mathbf{F}_\text{a-i},
\end{align}
where
\begin{align}
	\mathbf{F}_{\text{trap}} / m_\text{ion} = -\frac{\Omega_\text{rf}^2}{4}\left\lbrace \mathbf{a} + 2\mathbf{q} \cos\left(\Omega_\text{rf}t +
	\phi_\text{rf} \right)\right\rbrace \mathbf{r}_\text{ion}
\end{align}
is the force of the Paul trap with driving frequency $\Omega_\text{rf}/2\pi$ and phase $\phi_\text{rf}$, stability parameters $\mathbf{a}$ and $\mathbf{q}$, and the mass $m_\text{ion}$ and position $\mathbf{r}_\text{ion}$ of the ion~\cite{leibfried2003quantum}. We model the trap parameters after the Freiburg setup (for a detailed list see Tab.~\ref{tab:simulation_parameters}), denoting the two radial directions as $x$ and $y$ and the axial direction as $z$. To account for the finite size of the trap we calibrated $q_z$ and found it to be on the order of a few percent of $q_{x/y}$, which leads to a small axial micromotion. The next contributing term
\begin{align}
	\mathbf{F}_{\text{dc}} = e \,\bm{\mathcal{E}}_\text{disp}
\end{align}
is the force due to externally applied displacement fields $\mathbf{\mathcal{E}}_\text{disp}$. Finally,
\begin{align}
	\mathbf{F}_{\text{a-i}} & = \left( \frac{4 C_4}{\left\lVert\mathbf{r}_\text{a-i}\right\rVert^5} - \frac{6
		C_6}{\left\lVert\mathbf{r}_\text{a-i}\right\rVert^7} \right)
	\frac{\mathbf{r}_{\text{a-i}}}{\left\lVert\mathbf{r}_\text{a-i}\right\rVert}
\end{align}
is due to the atom-ion interaction potential, with the van-der-Waals coefficient $C_6$ and the relative atom-ion position $\mathbf{r}_\text{a-i}$. We phenomenologically choose $C_6$, such that the classical turning point of the combined potential matches that obtained from coupled-cluster calculations for the ground-state singlet $^1\Sigma^+$ channel~\cite{weckesserObservationFeshbachResonances2021}. As the confinement of the atoms is typically much weaker than that of the ions, we consider them free with $\mathbf{F}_\text{atom}=-\mathbf{F}_\text{a-i}$.

To model the effect of elastic collisions of the ion with thermal atoms, we follow~\cite{Fuerst2018}. In short, we initially position the ion at its displaced equilibrium position with zero kinetic energy. We then randomly place an atom on a sphere with radius $r_\text{init}$ around the ion whose tangential velocities are drawn from Gaussian distributions with $\sigma=\sqrt{k_\text{B} T_\text{atom} / m_\text{atom}}$ and normal velocity drawn from a Weibull distribution with shape $k=2$ and scale $\lambda = \sqrt{2k_\text{B}T_\text{atom}/m_\text{atom}}$. The eom are integrated until $|\mathbf{r}_\text{a-i}|>1.01\, r_\text{init}$, at which point a new thermal atom is seeded on a sphere around the current ion position. To control the collision energy, displacement fields of different magnitude $|\bm{\mathcal{E}}_\text{disp}|$ are applied in the $x$-$y$-plane. For each value of the displacement field, we simulate $N_\text{traj}=200$ trajectories with $N_\text{col}=5000$ collisions each. When randomly drawing ion velocities from the resulting trajectories, we ensure that the median collision energy has settled to be robust against equilibration dynamics. To integrate the eom we use a 12th order adaptive Runge-Kutta-Nyström solver~\cite{Rackauckas2017} with relative tolerance $10^{-16}$ and a forced maximum timestep of $\Delta t =2\pi/(100\,\Omega_\text{rf})$.

\begin{table}[b]
	\caption{\textbf{Parameters used for the numerical simulation of atom-ion collisions.} The displacement field was applied along $(\mathbf{e}_x+\mathbf{e}_y)/\sqrt{2}$. The parameters are modeled after the Freiburg hybrid setup for Li-Ba$^+$.}
	\centering
	\begin{tabular}{c c c}
		\toprule
		Parameter                   & Value                      & Description                                \\
		\midrule
		$\Omega_\text{rf}$          & $2\pi\times\SI{1.43}{MHz}$ & rf angular frequency                       \\
		$\phi_\text{rf}$            & randomized                 & rf drive phase                             \\
		$a_x$                       & 0                          & dc stab. param.                            \\
		$a_y$                       & \num{-1.04e-4}             & dc stab. param.                            \\
		$a_z$                       & $-a_y$                     & dc stab. param.                            \\
		$q_x$                       & \num{-0.1309}              & rf stab. param.                            \\
		$q_y$                       & \num{0.1274}               & rf stab. param.                            \\
		$q_z$                       & \num{0.0035}               & rf stab. param.                            \\
		$|\mathcal{E}_\text{disp}|$ & $3\times10^{-3}$--3\,V/m   & displacement field                         \\
		$m_\text{ion}$              & \SI{138}{u}                & ion mass                                   \\
		$m_\text{atom}$             & \SI{6}{u}                  & atom mass                                  \\
		$C_4$                       & \SI{2.803e-57}{Jm^4}       & pol. coefficient                           \\
		$C_6$                       & \SI{1.954e-76}{Jm^6}       & vdW coefficient                            \\
		$T_\text{atom}$             & \SI{0.7}{\micro\kelvin}    & atom temperature                           \\
		$r_\text{init}$             & \SI{0.9}{\micro\meter}     & atom seed radius                           \\
		$N_\text{col}$              & 5000                       & coll. per trajectory                       \\
		$N_\text{traj}$             & 200                        & trajectories per $\mathcal{E}_\text{disp}$ \\
		\bottomrule
	\end{tabular}
	\label{tab:simulation_parameters}
\end{table}

\end{document}